\documentclass[reprint,superscriptaddress]{revtex4-2}
\usepackage[T1]{fontenc}
\usepackage[utf8]{inputenc}
\usepackage{url}
\usepackage{amsmath}
\usepackage{mathtools}
\usepackage{bbold}
\usepackage{mathrsfs}
\DeclarePairedDelimiter{\abs}{\lvert}{\rvert}
\usepackage{graphicx}
\usepackage{epstopdf} 
\usepackage{wrapfig}
\usepackage{braket}
\usepackage{xcolor}
\usepackage{float}

\begin{document}

\newcommand{\giovanni}[1]{{\color{blue}#1}}

\author{Giovanni Pecci}
\affiliation{Univ. Grenoble Alpes, CNRS, LPMMC, 38000 Grenoble, France}

\author{Patrizia Vignolo}
\affiliation{Université Côte d’Azur, CNRS, Institut de Physique de Nice, 06560 Valbonne, France}

\author{Anna Minguzzi}
\affiliation{Univ. Grenoble Alpes, CNRS, LPMMC, 38000 Grenoble, France}

\title{Universal  spin mixing oscillations in a strongly interacting  one-dimensional Fermi gas}

\begin{abstract}
We study the spin-mixing  dynamics of a one-dimensional  strongly repulsive Fermi gas  under harmonic confinement. By employing a mapping onto an inhomogeneous isotropic Heisenberg model and the symmetries under particle exchange, we follow the dynamics till very long times. Starting from  an initial spin-separated state, we observe superdiffusion,  spin-dipolar large amplitude oscillations and thermalization. We report  a universal scaling of the oscillations with particle number $N^{1/4}$. Our study puts forward one-dimensional correlated fermions as a new system to observe the emergence of non-equilibrium universal features. 
\end{abstract}

\maketitle

Elucidating the dynamics of interacting Fermi gases is important for understanding a large variety of physical phenomena, from condensed matter  to plasmas and astrophysical objects as neutron stars. The strongly out-of-equilibrium dynamics of interacting quantum systems is currently one of the most challenging open problems. 

In this context, the spin dynamics deserves  a specific focus. Spin currents can be easily damped by inter-particle collisions \cite{PhysRevLett.98.266403} and the  continuity equation for the  spin density includes both orbital current and spin torque contributions \cite{RALPH20081190}. Spin drag is another manifestation of interactions among the spin species, inducing spin-diffusive or non-dissipative dynamics depending on the interaction regimes \cite{PhysRevB.62.4853,PhysRevB.65.085109,PhysRevLett.98.266403,Enss2012,Carlini2021,valtolina2017exploring}.
Ultracold atomic gases provide an ideal platform for exploring in isolated conditions the out-of-equilibrium spin dynamics \cite{PhysRevA.91.023620,barfknecht2019dynamics,PhysRevB.99.014305}. In a three dimensional geometry, the oscillatory dynamics of a strongly interacting Fermi gas with  initially spatially separated spin components was studied in \cite{sommer2011universal}. The spin drag, spin diffusivity and spin susceptibility   were obtained, and  a universal limit for spin diffusivity at low temperature was reported for the unitary Fermi gas. 

A relevant question is what happens to the above quantities when reducing the dimensionality of the system to quasi one-dimensional, and what type of universality emerges.  One-dimensional (1D) systems display specific features, as the enhancement of quantum fluctuations and  correlations and they can be described by a wealth of theoretical and numerical  methods \cite{gaudin1967systeme,PhysRevLett.19.1312,PhysRevA.105.013314,koutentakis2019probing}. The quantum dynamics may be strongly affected by the geometrical constraints, as well as by the presence of a large number of conserved quantities, as demonstrated e.g.~in the quantum Newton's cradle experiment \cite{kinoshita2006quantum}.

We address this question by following the dynamics of strongly repulsive fermions subjected to a longitudinal harmonic confinement in a tight waveguide. As in the three-dimensional case of  Ref.\cite{sommer2011universal}, we start from an initially imbalanced state with all spin up on the left and all spin down on the right of the harmonic trap, and we follow the damped oscillations of the magnetization. While the fully quantum dynamics at arbitrary interactions can be followed only at short times with a classical simulator, we focus here on the strongly correlated regime of very large interactions, close to the integrable point at infinite repulsions \cite{PhysRevX.7.041001,volosniev2014strongly,PhysRevLett.100.160405,Decamp_2016,PhysRevA.84.023626,PhysRevLett.115.215301,doi:10.1126/science.aag1635}. In this regime, the dynamics of the charge and spin decouple, and the spin dynamics can be followed exactly till very long times by means of a mapping onto the one of an inhomogeneous, isotropic Heisenberg model \cite{sutherland2004beautiful,franchini2017introduction} with site-dependent couplings \cite{PhysRevA.90.013611,PhysRevA.94.023606}.  

\begin{figure}[H]
\includegraphics[width=1\columnwidth]{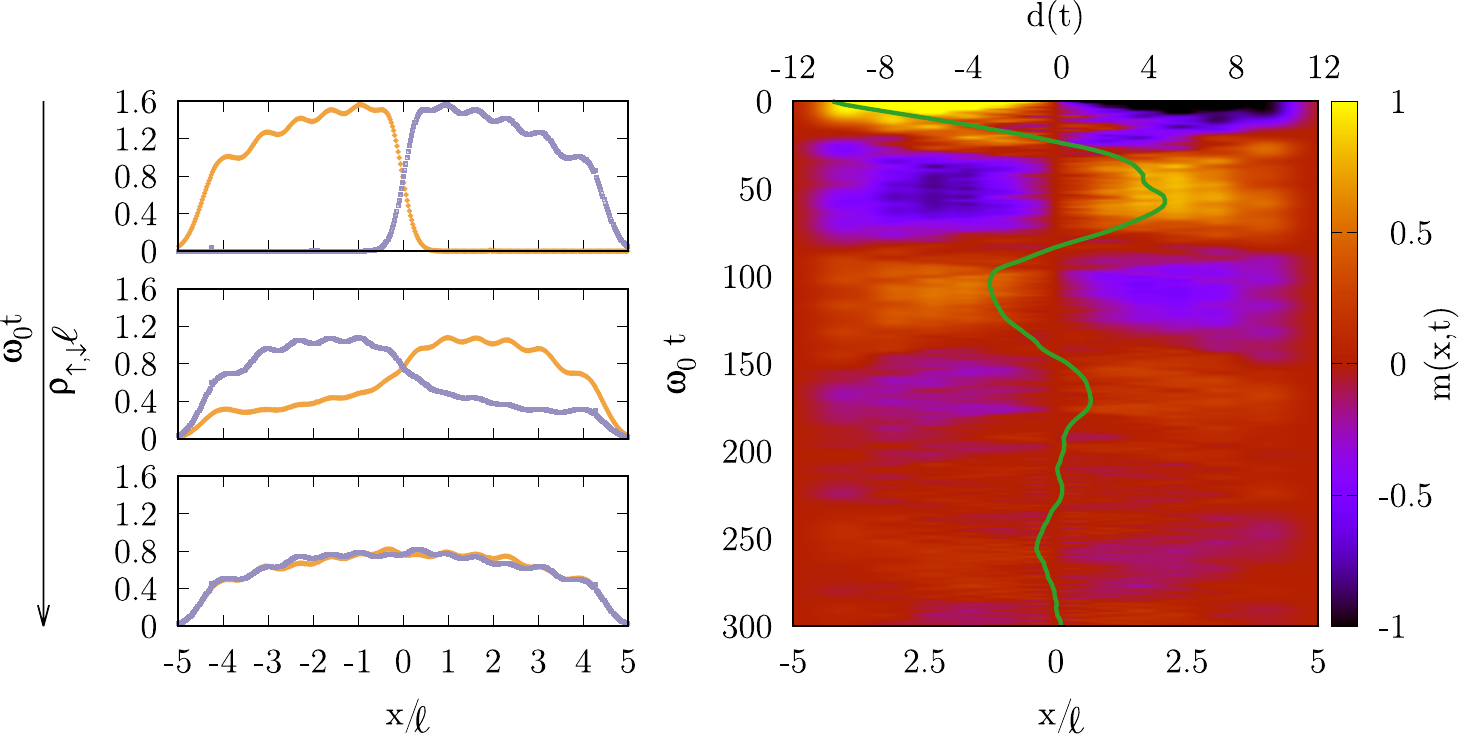}
\vspace{-0.5cm}
\caption{Left panel: spin up $\rho_\uparrow$ in orange (light grey) and down $\rho_\downarrow$ in violet (dark grey) spatial  densities (in units of the inverse harmonic oscillator length $\ell^-1$, with  $\ell=\sqrt{\hbar/m\omega_0}$) as a function of position in the trap (in units of $\ell$)  at times $\omega_0t=0, 33, 200$ from top to bottom. The two  initially separated clouds start oscillating in the trap and eventually fully mix, approaching  to a zero-magnetization state.  Right panel: magnetization as a function of $x$ (in units of  $\ell$)  and $t$ (in units of $\omega_0^{-1}$) for $N=12$ fermions. The green line corresponds to center of mass $d(t)$ of the magnetization.}
\label{Fig0}
\end{figure}

 An overview of the full spin dynamics is provided in Fig.~\ref{Fig0}, where three main dynamical regimes arise. At short times, we predict the emergence of a superdiffusive behaviour, compatible with Kardar-Parisi-Zhang (KPZ) universality, in striking difference from the diffusive one found in the three-dimensional counterpart\cite{sommer2011universal}. We thus identify  1D correlated fermions as a new system to observe  the emergence of non-equilibrium universality, largely explored in   homogeneous  Heisenberg models \cite{ljubotina2017spin,PhysRevE.71.036102,PhysRevLett.121.230602,PhysRevLett.123.186601,PhysRevLett.122.210602,PhysRevLett.122.127202,PhysRevB.101.121106,Bastianello_2022} and experimentally  evidenced in quantum magnets  and in ultracold atoms on a lattice \cite{PhysRevB.101.041411,Iversen_2020,wei2021quantum,
 scheie2021detection}.
 At intermediate times, we observe large-amplitude spin-dipole oscillations and we obtain the spin drag decay rate. We unveil a  $N^{1/4}$ scaling in the oscillation frequency, implying a slow-down of the motion and the decrease of the zero-temperature spin drag rate as the particle number grows. 
 
 At long times, the oscillations are damped out and the system thermalizes to the diagonal ensemble   \cite{rigol2008thermalization}. From the analysis of the energy levels distribution we find that the system is weakly non-integrable. The proposed setup allows to explore the conditions for emergence of non-equilibrium universal behaviour in relation to the breaking of its integrability in one dimension. 

\paragraph*{Model and dynamics}
We consider a one-dimensional SU(2) interacting Fermi gas confined in a harmonic trap. The Hamiltonian for such system reads
\begin{equation}
H = \sum_{i=1}^N \Bigl( \frac{p_i^2}{2m} + \frac{m\omega_0^2 x_i^2}{2} \Bigr) + g \sum_{i \neq j}\delta(x_i-x_j),
\label{hamiltonian}
\end{equation}
where $N = N_\uparrow+N_\downarrow$ is the total number of particles and we take $N_\uparrow=N_\downarrow$, $\omega_0$ is the frequency of the harmonic trap and we model the interspecies interaction using a delta potential of strength $g$.  Hamiltonian (\ref{hamiltonian}) is characterized by the symmetry under exchange of particles having the same spin. For SU(2) fermions, the eigenstates can be classified by the irreducible representations of the permutation group (see eg.~\cite{Decamp_2016}).

We focus on the strongly repulsive limit $g \to \infty$: in this regime the model is exactly solvable \cite{volosniev2014strongly} and the wavefunction is given by:
\begin{equation}
\Psi =\sum_P \theta(x_{P(1)} < ... < x_{P(N)}) \ a_P \Psi_A (x_1, ...x_N),
\label{wavefunction}
\end{equation}
where $\theta(x_1 < x_2 < x_3 <... x_N)$ is the characteristic function of the coordinate sector $\{x_1 < x_2 < x_3 < ... <x_N \}$, $a_P$ are phases depending on the spin ordering of the corresponding coordinate sector and the summation is performed over all the possible permutations $P$ of $N$ elements. The function $\Psi_A$ is the wavefunction of a $N$-particle non-interacting Fermi gas in the same external potential, i.e. the anti-symmetric product of $N$ eigenfunctions of the harmonic oscillator. Remarkably, in the $g\rightarrow \infty$ limit the spin and spatial ('charge') degrees of freedom are decoupled in the wavefunction. 

We determine the phases $a_p$ in Eq.(\ref{wavefunction}) to first order in $1/g$ by mapping the Hamiltonian (\ref{hamiltonian}) into an effective spin chain: 
\begin{equation}
H_s = (E_F - \sum_i^{N-1} J_i)\mathbb{1} + \sum_{i=1}^{N-1} J_i P_{i,i+1},  
\label{Spin_H}
\end{equation}
where $P_{i,i+1}$ is the transposition operator on the chain of $N$ sites and $E_F=N^2\hbar\omega_0/2$ is the Fermi energy. The coefficients $J_i$ are site-dependent hopping parameters of the chain, carrying information on the external potential and on the atom-atom interaction of the original fermionic problem~\cite{supmat}. The explicit expression reads ~\cite{conan}:
\begin{equation}
J_i = \frac{1}{g} \int_{-\infty}^{\infty} dx_1...dx_N ~\delta(x_i - x_{i+1}) \ \theta(x_1 < ... < x_N) \Bigl\lvert \frac{\partial \Psi_A}{\partial x_i}\Big\lvert ^2.
\label{J}
\end{equation}

We classify the basis vectors of the Hilbert space associated to (\ref{Spin_H}) according to the spin ordering on the chain (the so-called snippet basis \cite{Deuretzbacher2008}). For example, for $N_\uparrow = N_\downarrow =2$ the vector $\ket{\uparrow \uparrow \downarrow \downarrow}$ indicates that all the spins $\uparrow$ are placed in the left half of the chain.  Therefore, the dimension of the Hilbert space is $s =\frac{N!}{N_\uparrow! N_\downarrow!}$. 
The diagonalization of Eq.~(\ref{Spin_H}) allows us to calculate the $a_P$ and thus several observables such as the spin densities $\rho_{\uparrow,\downarrow}(x,t)$. This allows to study the dynamics of the trapped system with an arbitrary initial state.

In this work we follow the fermion dynamics starting from the initially strongly out-of-equilibrium  state $|\chi(t=0)\rangle=\ket{\uparrow \uparrow \uparrow ... \downarrow \downarrow \downarrow}$, as in Ref.\cite{sommer2011universal}, where the spins up and down are separated in the two opposite sides of the trap.  
Since the harmonic trap is unchanged, the spatial part of the wavefunction (\ref{wavefunction}) is constant during the motion, hence  $J_i$ are constant in time. The  time evolution involves only the spin degrees of freedom and can be obtained using the effective spin chain Hamiltonian (\ref{Spin_H}). Recalling that the spin operators are related to the permutation operator by the relation $P_{k,k+1} = \frac{1}{2} (\mathbb{1} + \sigma_k \cdot \sigma_{k+1})$, Hamiltonian (\ref{hamiltonian}) can be mapped to the one of an inhomogeneous isotropic Heisenberg model $H_H=\sum_{j=1}^{N-1} J_j \vec \sigma_j \cdot \vec \sigma_{j+1}$, but {\em in particle space}, ie each lattice site is associated to a particle index.
The equation of motion for the spin operator $\vec{S}_j = \frac{1}{2}(\sigma_j^x, \sigma_j^y, \sigma_j^z)$ for the $j$-th particle reads
\begin{equation}
\frac{dS_j^{\mu}}{dt} = i[H_s,S_j^{\mu}] = (\vec{\tau}_j \times \vec{S}_j)^{\mu} ,
\label{continuity equation}
\end{equation}
where $\mu = x,y,z$.
and  $\vec{\tau}_j=J_{j-1}\vec{\sigma}_{j-1} + J_j \vec{\sigma}_{j+1}$ is the torque acting on a fixed particle due to the coupling with the neighbouring ones \cite{supmat}. 
As a result of Eq.~(\ref{continuity equation}) we conclude that in our case the spin dynamics is entirely due to spin torque ~\cite{supmat}.  

The experimentally accessible component of such spin vector is $S_j^z$, associated to the local magnetization 
\begin{equation}
m(x,t)=\sum_{j=1}^N  m_j(t)\rho_j(x),
\end{equation}
where 
$\rho_j(x)$ is the spatial density of the $j$-th particle in the trap \cite{PhysRevA.90.013611,supmat}, $m_j(t) = \langle \chi(t)| S_{j}^z|\chi(t)\rangle$ and $|\chi(t)\rangle = e^{-i H_s t} |\chi(0)\rangle$ is the time-evolved spin state, obtained from the diagonalization of $H_s$ by exploiting all its symmetries. The magnetization is experimentally accessible by recording the population imbalance among $\uparrow$ and $\downarrow$ fermions, $m(x,t) = \rho_\uparrow(x,t)-\rho_\downarrow(x,t)$. 

Another important observable for the dynamics is the spin current density 
\begin{equation}
j(x,t) = \frac{1}{2}\sum_{j=1}^{N-1} j_j(t) (\rho_j(x) + \rho_{j+1}(x)),
\label{spincurrent}
\end{equation}
where $j_j(t)$ are obtained from the $z$ component of Eq.~(\ref{continuity equation}), 
$j_j(t) = J_j (\sigma_j^x \sigma_{j+1}^y - \sigma_j^y \sigma_{j+1}^x ).$
We detail in the following the spin-mixing  dynamics in the various relevant time regimes. Systems up to $N=12$ particles have been analyzed using exact diagonalization, while we used truncated Taylor series approximation of the time evolution operator \cite{al2011computing} and tDMRG \cite{PhysRevLett.108.245302} to study larger systems.

\begin{figure}[htb]
\includegraphics[width=1\columnwidth]{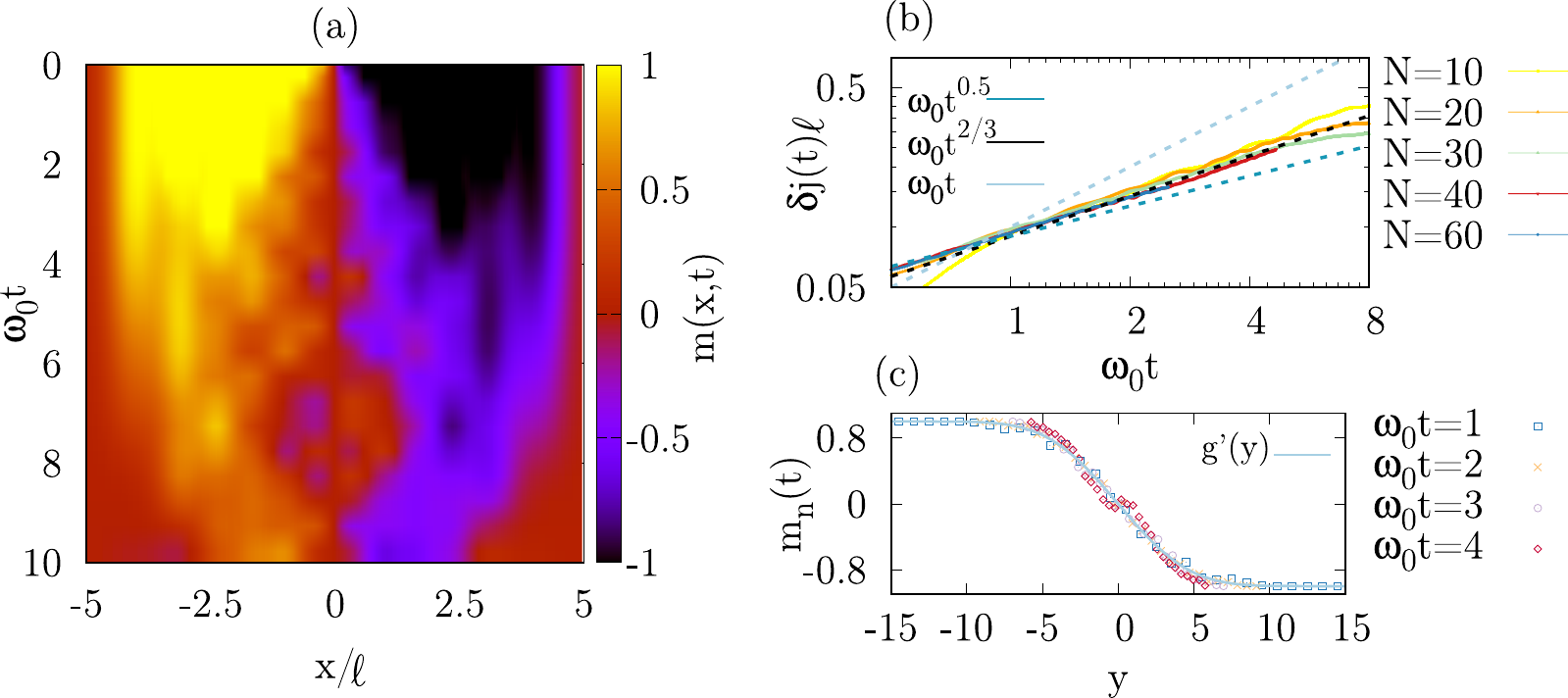}
\vspace{-0.7cm}
\caption{
(a) Early-time magnetization dynamics $m(x,t)$ from exact solution for $N=12$  as a function of space ($x$ in units of $\ell$) and time $t$ (in units of $\omega_0^{-1}$). (b) Integrated current $\delta j(t)$ in units of $\ell^{-1}$ from DMRG, as a function of time $t$ (in units of $\omega_0^{-1}$) evaluated in the center of the trap $x=0$ (solid lines), together with ballistic $\delta j \sim  t$ , KPZ $ \delta j \sim t^{2/3}$ and  diffusive behaviour $ \delta j \sim t^{1/2}$ (dashed straight lines). Larger number of particles are associated with increasingly darker colors. (c)  KPZ scaling of the magnetization $m_n(t)$, shown  as a function of  $y =n/(\omega_0 t)^{2/3}$. We show the comparison with the derivative of the suitably renormalized 1d KPZ scaling function $g(y)$~\cite{kpz}.}
\label{Short}
\end{figure}

\paragraph*{Short times}
The short time dynamics, before the first spin oscillation,  is shown in Fig.~\ref{Short}.  We  show the magnetization as a function of space and time, showing that the initially sharp magnetization interface spreads with time, till it starts feeling the effect of the confining potential. 
To identify the nature of the magnetization spreading, it is useful to study  the time-integrated spin current  density as in \cite{ljubotina2017spin},
$\delta j(t) = \int_{0}^t dt' ~ j(0,t')$, where $j(x,t)$ is defined in Eq.(\ref{spincurrent}) and it is calculated at the center of the trap. We see that the integrated spin current displays a superdiffusive behaviour $\delta j(t)\sim t^\eta$, with power law exponent $\eta\sim 0.638(1)$. This is compatible with the  value $\eta=2/3$ predicted for the homogeneous spin chain and clearly not ballistic nor diffusive.Remarkably, low-energy dynamics described by Luttinger liquid predicts ballistic behaviour \cite{PhysRevLett.98.266403}: the deviation from this prediction discloses the marked out-of-equilibrium features of the physical system that are described in an exact way by our model. Using DMRG calculations we have checked that the KPZ region persists for larger numbers of particles (see Fig.\ref{Short}) The deviation at later times from KPZ behaviour is due to onset of the oscillatory dynamics associated to the presence of the external trap. Within the KPZ region, we find that the magnetization profiles collapse onto each other if plotted as a function of $x_n/(\omega_0 t)^{1/z}$,  $z=3/2$ being the KPZ value for the dynamical critical exponent.

\begin{figure*}[t]
\includegraphics[width=1\textwidth]{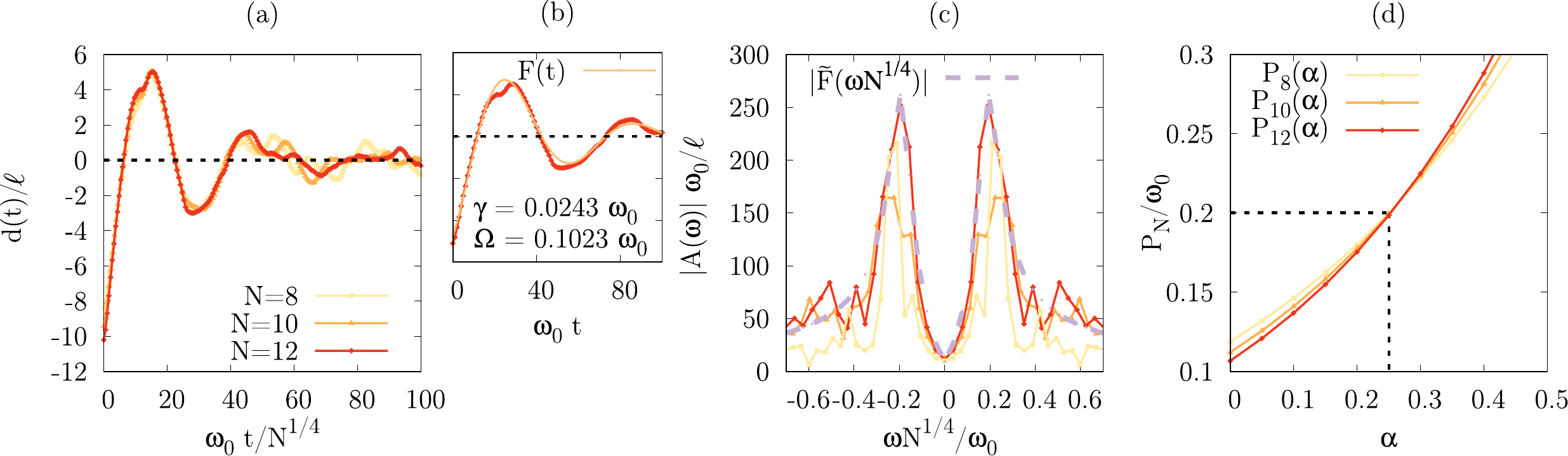}
\caption{(a) Center of mass of the magnetization $d(t)$, in units of $\ell$,   as a function of time,  in units of $\omega_0^{-1}$, and  scaled by a factor $N^{1/4}$ to evidence the universal behaviour of the oscillations. (b) $d(t)$ for $N=12$ fitted with a damped harmonic oscillator $F(t) =f_0 e^{-\gamma t}\cos(\Omega t)$. (c) Modulus of $A(\omega)$, in units of $\ell/\omega_0$, for different number of particles as a function of the universal frequencies $\omega N^{1/4}/\omega_0$, compared to
the modulus of the Fourier transform $\tilde{F}(\omega)$ of the fit $F(t)$ (dotted violet line). Colors codes are the same as in panel (a). (d) Position $P_N(\alpha)$ of the peaks of $\Tilde{F}(\omega N^\alpha)$, in units of $\omega_0$, as a function of the scaling parameter $\alpha$. }
\label{Intermediate}
\end{figure*}

\paragraph*{Intermediate times} 
We next focus on the intermediate time regime, when the particles undergo large-amplitude spin-dipole oscillations in the trap. We follow the  center of mass oscillations of the magnetization, 
\begin{equation}
d(t) = \frac{1}{N}\int_{-\infty}^{\infty} dx \ x \ m(x,t).
\label{doft}
\end{equation}
The time evolution of $d(t)$ is shown in Fig.~\ref{Intermediate}a for various values of the number of particles. We observe damped oscillations tending to a plateau corresponding to zero magnetization.  At later times (not shown in the figure), the dynamics undergoes several partial revivals, as expected since we describe a closed quantum system. 
Quite remarkably, we find  that the various curves for different particle numbers collapse one to another if we scale the time axis by a factor $N^{\alpha}$ with $\alpha=0.25$. Using DMRG simulations up to $N=60$ particles \cite{supmat}, we have tested that the scaling is robust at increasing particle numbers. The magnetization oscillations are well approximated by  a damped harmonic oscillator $F(t) = f_0 e^{-\gamma t} \cos(\Omega t)$(see Fig.~\ref{Intermediate}b). This allows us to obtain  the oscillation frequency $\Omega$ for the various $N$ values. 
We find that the damping rate $\gamma$  doesn't depend on the number of particles. Combining the two values, we obtain the
 spin drag rate as $\Gamma_{sd}=\Omega^2/\gamma$  \cite{PhysRevB.62.4853,PhysRevB.65.085109,sommer2011universal,spindrag}.

We also perform a spectral analysis of $d(t)$ by introducing the spectral function
$A(\omega) = \int_{0}^{\infty}dt\ d(t) e^{i\omega t}$.
In Fig.~\ref{Intermediate}c we show $\abs{A(\omega)}$ and the Fourier transform of the fitted signal $\Tilde{F}(\omega) = \int_{0}^\infty dt\ F(t) e^{i\omega t}$ as a function of the rescaled frequencies. The spectral function shows two main peaks centered around $\pm \Omega_{univ}$.  Several excitation frequencies contribute to the overall shape and the linewidth of $\Tilde{F}(\omega)$ ~\cite{supmat}. To estimate the scaling exponent $\alpha$ we evaluate the position  $P_N(\alpha) = \omega_{P}N^\alpha$ of the maximum of $\Tilde{F}(\omega N^{\alpha})$ at positive frequencies, such that $\Tilde{F}(\omega_{P} N^{\alpha})= \max_{\omega>0}\Tilde{F}(\omega N^{\alpha})$,
as a function of a scaling exponent $\alpha$. As we show in Fig.~\ref{Intermediate}d, the universal scaling is reached for $\alpha = 0.25$.

The universal scaling observed in Fig.~\ref{Intermediate} allows us to estimate the spin-dipole oscillation frequency at larger $N$ as $\Omega_N\simeq \Omega_{univ}/ N^{1/4}$, with $\Omega_{univ}\simeq 0.19 ~\omega_0$. Correspondingly, we find that 
 the spin drag scales as $\Gamma_{sd}=\Omega_{univ}^2/(\gamma N^{1/2})$, hence vanishing  at large particle numbers, as also predicted in \cite{PhysRevLett.98.266403} for low-energy excitations of the spectrum.
 
\begin{figure}
\includegraphics[width=1\columnwidth]{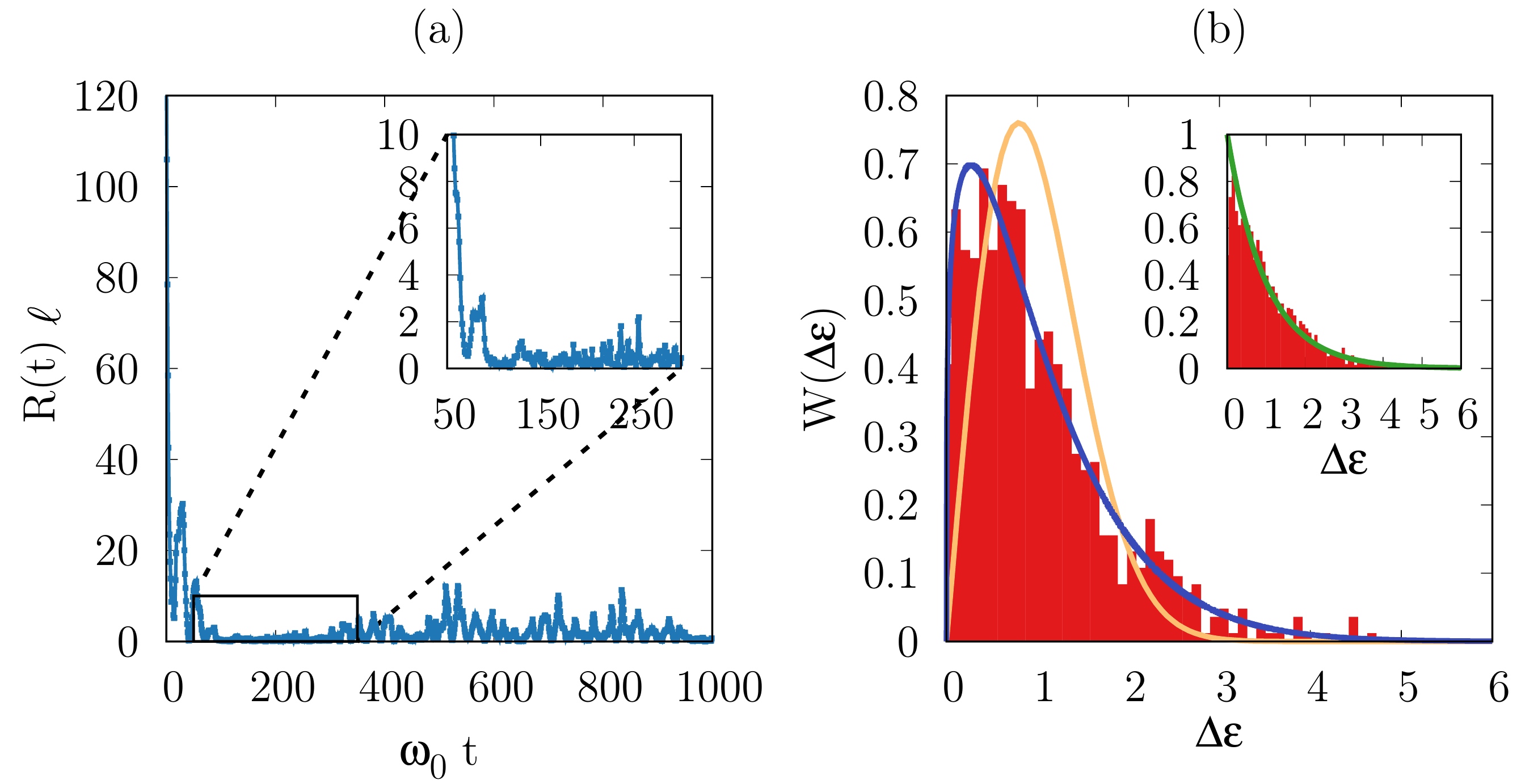}
\vspace{-0.5cm}
\caption{(a) Distance $R(t)$ (in units of $\ell^{-1}$) as a function of time (in units of $\omega_0^{-1}$).  The inset shows a zoom of the area indicated by the rectangle. (b) Level spacing distribution  $W(\Delta \epsilon)$ for the unfolded spectrum in a sector at fixed symmetry. The orange and the blue curves show respectively the Wigner-Dyson $W_{WD}(\Delta \epsilon)$ and the Brody distribution $W_B(\Delta \epsilon)$ with  $\beta = 0.22$. The inset shows the level-spacing distribution of the whole unfolded spectrum and the Poisson distribution $W_P(\Delta \epsilon)$ (green line). In all the panels, $N=14$.
\vspace{-0.3cm}}
\label{Long}
\end{figure}

\paragraph*{Long times}
Finally, we study the long-time regime at which the damped dynamics becomes dominant and the system approaches to a zero-magnetization state. Since the Hamiltonian (\ref{hamiltonian}) is not integrable at finite interaction strength, we expect some traces of chaoticity to emerge during the dynamics \cite{rigol2008thermalization,ueda2020quantum,PhysRevE.85.036209}. In this case the system  thermalizes to a state described by the diagonal ensemble, coinciding in our case  with the microcanonical ensemble~\cite{quantumchaosreview}. We verify this by calculating  the distance  $R(t) = \int dx~ \abs{\rho_{\uparrow}(x,t) - \rho_{\uparrow,MC}(x)}^2$ between the spin up density and its value in the microcanonical ensemble $\rho_{\uparrow,MC}(x)$. The results are presented in 
Fig.~\ref{Long}a. At times corresponding to the zero-magnetization plateau in Fig.~\ref{Long},  $R(t)$ vanishes and the spin density approaches to the steady state value. At later times, revivals occur and the system deviates from this configuration.

To further provide evidence for chaotic behaviour, we analyze the level-spacing distribution $W(\Delta \epsilon)$\cite{GUHR1998189,PhysRevE.81.036206,BORGONOVI20161}, constructed using  the unfolded dimensionless energy levels~\cite{gubin2012quantum,PhysRevE.85.036209,unfolded,gubin2012quantum,santos2009transport}. 
The spectrum of an integrable system follows a Poissonian distribution $W_P(\Delta \epsilon)\!\!=\!\!e^{-\Delta \epsilon}$, while a chaotic system is described by a Wigner-Dyson one $W_{WD}(\Delta \epsilon)\!\!=\!\!\frac{\pi}{2} \Delta \epsilon ~e^{-\frac{\pi}{4}\Delta \epsilon^2}$\!. 
We interpolate between the two regimes, thus quantifying the level of chaoticity encoded in the spectrum, through the Brody distribution~\cite{RevModPhys.53.385}:
\begin{equation}
W_B(\Delta \epsilon) =(\beta + 1)b\Delta \epsilon^\beta e^{-b\Delta \epsilon^{\beta+1}},
\label{BrodyDistr}
\end{equation}
where $ \quad b = \{\Gamma[(\beta+2)/(\beta+1)]\}^{\beta+1}$ and $\Gamma$ is the Euler Gamma function.  The  Brody distribution reduces to the Poisson or  Wigner-Dyson ones for $\beta=0$ or $1$ respectively.

To obtain the level-spacing distribution it is important to take into account  the symmetries of the system ~\cite{Stone_2010}, which in our case are the spatial parity and the symmetry under particle exchange. Our choice of basis vectors allows us to readily check the parity of the eigenstates. In order to identify the symmetry under particle exchange associated to a given Young tableau, we diagonalize the Heisenberg Hamiltonian in the basis of the permutational symmetry~\cite{PhysRevLett.113.127204,Decamp_2016}. We then  partition the energy levels according to the quantum numbers of the corresponding eigenstates. 
In the inset in Fig.\ref{Long}b we show the distribution of all the unfolded level spacings, irrespectively of the symmetry constraints. In this case the chaoticity is hidden and the distribution is Poissonian. The level-spacing distribution  of the largest  subspace at fixed symmetry is shown in the main panel of  Fig.\ref{Long}b. We find that the  level-spacing distribution is well described by the Brody distribution with parameter $\beta = 0.22$. This shows that large interactions destroy only partially the integrability of the infinite-repulsion model. A moderately chaotic  behaviour also emerges from the study of the localization properties of the eigenstates of the Hamiltonian (\ref{hamiltonian})  ~\cite{supmat}. Such intermediate regime is typical of integrable systems subjected to small perturbations \cite{PhysRevE.85.036209,Santos_2004,PhysRevE.81.036206}.

\paragraph*{Conclusions}
We have studied the strongly  out-of-equilibrium spin-mixing dynamics of  repulsive 1D fermions under harmonic confinement, starting from  an initial spatially separated spin configuration. Thanks to the mapping to an inhomogeneous Heisenberg model on an effective lattice in particle space, we have followed the real-space magnetization dynamics till very long times. 
At short times, as specific of one-dimensional systems and different from the three-dimensional strongly interacting Fermi gas, we observe superdiffusive behaviour of the magnetization profile in time. The system here considered is weakly not integrable,hence equivalent to the case where KPZ universality was reported in the short times dynamics \cite{PhysRevLett.127.057201,Bastianello_2022,de2021subdiffusive}. Our observations call for the exploration of the universal properties of the corresponding spin model. 
At intermediate times, we have obtained damped spin-dipole oscillations characterized by a universal scaling of the oscillation time with $N^{1/4}$, thus predicting a slow-down of the oscillation and decrease of spin drag at large particle numbers. At long times, we find that the system thermalizes to a diagonal ensemble state thanks to its moderately chaotic behaviour. All our conclusions hold exactly for strongly repulsive interactions to order $1/g$. A study of itinerant fermions at arbitrary interactions and long times remains an open challenge.
Our results show that harmonically trapped strongly interacting fermions are a promising platform for exploring the many facets of the non-equilibrium quantum dynamics.

\paragraph*{Acknowledgements}
We would like to thank Maxim Olshanii and Giacomo Roati for stimulating discussions, Pierre Nataf for his assistance in the early stages of this work and  Piero Naldesi for his significant help with DMRG calculations. We acknowledge funding from the ANR-21-CE47-0009 Quantum-SOPHA project.

\bibliography{Spin_dynamics.bib}

\begin{thebibliography}{64}
\bibitem [{con()}]{conan}%
  \BibitemOpen
  \href@noop {} {}\bibinfo {note} {To compute the hopping coefficients $J_i $
  we used the open source code CONAN \cite{LOFT2016171}.}\BibitemShut {Stop}%
\bibitem [{spi()}]{spindrag}%
  \BibitemOpen
  \href@noop {} {}\bibinfo {note} {Here we assume that the spin drag is
  constant over the time of the damped oscillations. The study of a larger
  system would be required to test this hypothesis.}\BibitemShut {Stop}%
\bibitem [{unf()}]{unfolded}%
  \BibitemOpen
  \href@noop {} {}\bibinfo {note} {Unfolding is obtained by normalizing the
  energy levels such that the mean level spacing is equal to one everywhere in
  the spectrum.}\BibitemShut {Stop}%
\end{thebibliography}

\onecolumngrid 
\newpage

\section*{Supplemental material for 'Universal spin mixing  oscillations in a strongly interacting one-dimensional Fermi gas'}

\subsection{Strongly repulsive regime}
We show here an intrinsic property of the strong-coupling expansion we are using to study the system: a variation of the    coupling constant $g$ to a new value $g'$  induces a scaling of all many-body energy levels $E_n$ to new values $E'_n$ such that $g E_n=g' E_n'$, without affecting the eigenvectors. This follows from the expression of $H_s$ and of the coefficient $J_i$ given in the main text and may be used as to identify  the strongly interacting regime.

The above property implies that the frequencies $\omega$ driving the dynamics, i.e. the energy spacings, scale accordingly. As a consequence, by  rescaling the time scale likewise, the dynamics doesn't depend on the actual value of the coupling constant $g$. 

\begin{figure}[h!!]
\includegraphics[width=0.4\textwidth]{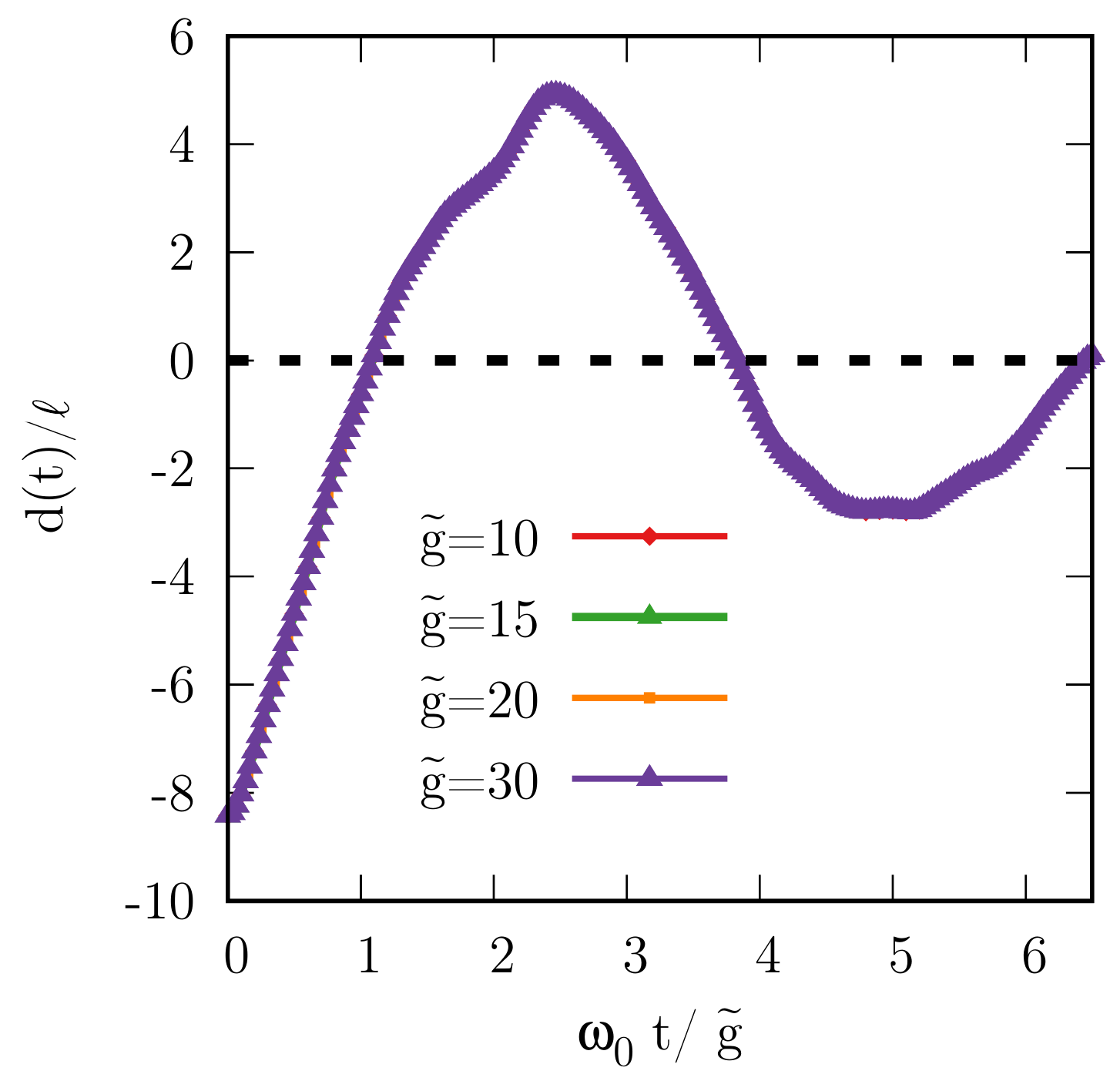}
\caption{Center of mass of the magnetization $d(t)$, in units of $\ell$, for $N=8$ particles,  as a function of the rescaled time $\omega_0 t/\tilde g$, for various values of the dimensionless coupling constant $\tilde{g}$. The curves corresponding to different $\tilde g$ are not resolved since they collapse one onto the other.}
\label{AppendixA}
\end{figure}

In Fig.~\ref{AppendixA} we show the center of mass of the magnetization $d(t)$ as a function of the time by rescaling the time axis of a factor of $1/\tilde{g}$, being $\tilde{g} = g/(\hbar \omega_0 \textit{l})$. We observe that, for various values of the coupling constant, the spin dynamics has exactly the same features.

\subsection{Continuity equation}
We demonstrate here Eq.~(5), showing that the dynamics we study in the main text is purely due to a torque moment among the spins. Consider indeed
\begin{equation}
[H_s, S_j^\mu] = \frac{1}{2} \sum_{l} \sum_{\lambda=x,y,z} J_l [\sigma_l^\lambda \sigma_{l+1}^\lambda, \sigma_j^\mu].   
\end{equation}
Decomposing the commutator and using the relation $[\sigma_i^\mu, \sigma_j^\lambda] =2i\sum_{\gamma} \delta_{ij} ~ \epsilon^{\mu \lambda \gamma}~ \sigma_i^\gamma$ we get 
\begin{equation}
\frac{dS_j^\mu}{dt} = -\sum_{\gamma, \lambda} \epsilon^{\mu\lambda\gamma} (J_{j-1}\sigma_{j-1}^\lambda + J_j \sigma_{j+1}^\lambda) \sigma^{\gamma}_j,
\label{torque}
\end{equation}
from which we readily obtain Eq.~(5).

In order to have a better understanding of the dynamics, we write the three components of the last equation on the snippet basis which is our computational basis. To do so, we recall the action of the Pauli matrices on the single-spin Hilbert space. From their explicit expression one sees that both $\sigma_j^x$ and $\sigma_j^y$ induce a spin flip on the $j$-th site, the latter also imprinting a spin-dependent phase. On the other hand, $\sigma_j^z$ doesn't invert the spin on the site $j$ and its action it's equivalent to the identity if the $j$-th spin is up and induces a phase shift of $\pi$ if this spin is down. Consequently, as the components $x$ and $y$ of Eq.~(\ref{torque}) involve respectively products of $\sigma_j^y \sigma_{j+1}^z$ and $\sigma_j^x \sigma_{j+1}^z$, they modify the total spin of the state. 

From the above considerations and since we are working at fixed total spin, we can assess that the expectation value of $\frac{dS_j^x}{dt}$ and $\frac{dS_j^y}{dt}$ are vanishing on the basis we are considering. This is not the case  for $\frac{dS_j^z}{dt}$, whose expectation value provides access to the expression of the spin current in terms of the permutation operator, 
\begin{equation}
\bra{e_p}j_l\ket{e_q} = 2J_l~(-1)^{\delta_{q(l),\uparrow}}  \bra{e_p}P_{k,k+1}\ket{e_q} (1 - \delta_{pq}).
\end{equation}

\subsection{Orbital current}
We show here that the particle orbital current is zero and consequently the dynamics described in the main text in the strongly interacting regime is only due to spin torque. 

The equation of motion for the density of spin $\uparrow$ particles reads (an analogous definition hold for spin $\downarrow$)
\begin{equation}
\frac{\partial n_{\uparrow}(x,t)}{\partial t} = \int_{-\infty}^{\infty} dx_1...dx_N ~\frac{\partial}{\partial t}\Bigl(\Psi^* \sum_{j=1}^{N_\uparrow} \delta(x-x_j)\Psi \Bigr).
\end{equation}
Using the wavefunction reported in Eq.(2) of the main text, and recalling that there is no time-dependence of the single-particle orbitals in the quench protocol here considered,  we obtain
\begin{equation}
\frac{\partial n_{\uparrow}(x,t)}{\partial t} = \sum_{P,Q} \Bigl( \frac{\partial a_P^*}{\partial t} a_Q + a_P^* \frac{\partial a_Q}{\partial t}\Bigr) 
\int_{-\infty}^{\infty}dx_1...dx_N~\sum_{j=1}^{N_\uparrow}\theta_P \theta_Q \delta(x-x_j) \abs{\Psi_A}^2,
\end{equation}
where we set $\theta_P = \theta(x_{P(1)}< x_{P(2)} < ... x_{P(N)})$. Remarkably, the product $\theta_P \theta_Q$ is non-zero only when the permutations $P$ and $Q$ coincide. Consequently, the expression above can be simplified as in the following:
\begin{equation}
\begin{split}
&\frac{\partial n_{\uparrow}(x,t)}{\partial t} = \\ 
&\sum_P \Bigl( \frac{\partial a_P^*}{\partial t} a_P + a_P^* \frac{\partial a_P}{\partial t}\Bigr) \sum_{j=1}^{N_\uparrow} \int_{-\infty}^{\infty}dx_1...dx_N \theta_P \delta(x-x_j) \abs{\Psi_A}^2 = \\ 
&\sum_P \Bigl( \frac{\partial a_P^*}{\partial t} a_P + a_P^* \frac{\partial a_P}{\partial t}\Bigr) \sum_{j=1}^{N_\uparrow}\rho_{P(j)}(x),
\label{this-eq-1}
\end{split}
\end{equation}
where in the last line we used the definition of the spatial density of the $j$-th particle in the coordinate sector individuated by the permutation $P$. From Eq. (\ref{this-eq-1}) we see that the only contribution to the equation of motion for the spin up density comes from the spin sector, whose time evolution is uniquely fixed by the Hamiltonian $H_s$. The total density ('charge') sector, that in the dynamical protocol we consider is constant in time, only contributes with a spatial envelope given by the densities of the particles in the harmonic trap.

\begin{figure}
\includegraphics[width=1\textwidth]{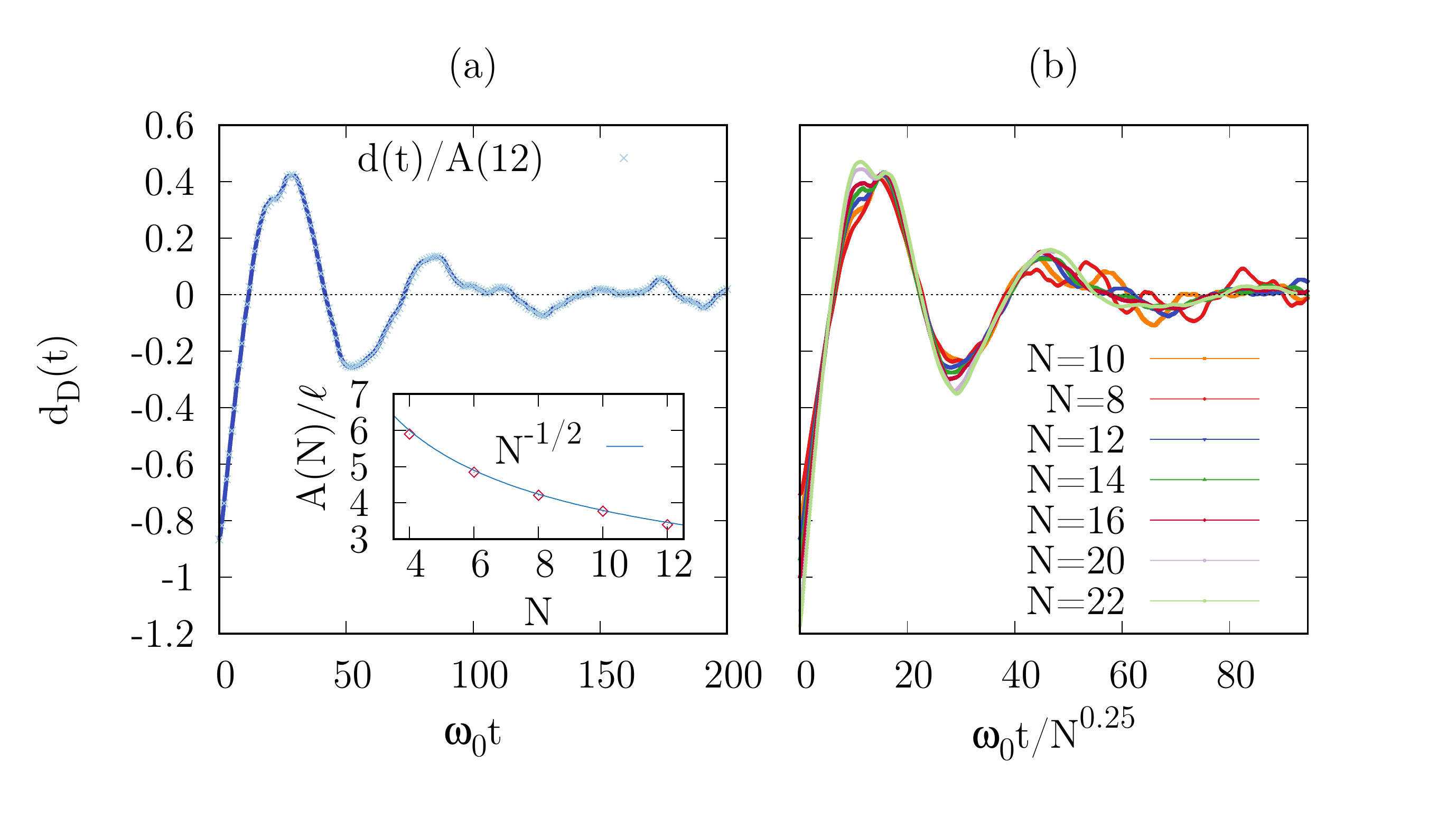}
\caption{(a) Discrete and continuum center of mass of magnetization for $N=12$ as a function of time. The two curves are proportional at any time. In the inset, we show the proportionality constant $A(N)$ as a function of $N$. (b) Discrete center of mass of magnetization $d_D(t)$ as a function of time for different number of particles. The time axis has been rescaled by a factor of $N^{1/4}$ to stress the universal behaviour of the frequency of the spin oscillations.}
\label{AppendixD}
\end{figure}

\subsection{Parity of oscillation frequencies}
We derive here the analytical expression for the magnetization, that illustrates the energy levels involved in the time evolution.  Starting from the definition given in the main text, the magnetization can be written as $m(x,t)=\sum_{j=1}^s m_j(t) \rho_j(x)$. The probability of the i-th particle to have magnetization $\pm 1$ is:
\begin{equation}
m_j(t) = \bra{\chi(t)}S_j^z\ket{\chi(t)} = \sum_{p=1}^s \abs{a_p(t)}^2(\delta_{p(j),\uparrow} - \delta_{p(j),\downarrow} ),
\label{mj}
\end{equation}
where $\ket{\chi(t)} = \sum_{p=1}^s a_p \ket{e_p}$ is the spin component of the wavefunction, being $\ket{e_p}$ the snippet basis. We also introduced $\delta_{p(j),\uparrow(\downarrow)}$ that is equal to one if the $j$-th component of the $p$-th element of the basis is a spin up (down) and zero otherwise. We want to study the symmetry of Eq.~(\ref{mj}), starting from the explicit expression for $\abs{a_p(t)}^2$,
\begin{equation}
\abs{a_p(t)}^2 = \sum_{l,k}^s e^{-i(E_k - E_l)t} f_{p,k}~f_{p,l}~\psi_{l0}~\psi_{k0},
\label{ap}
\end{equation}
where $f_{p,k} = \braket{e_p | \psi_k }$ and $\psi_{k0} = \braket{\psi_k|\chi(0)}$, with $\ket{\psi_k}$ eigenstates of $H_s$. When $N_\uparrow = N_\downarrow$ the Hamiltonian is symmetric under spin inversion: one can check that this implies $f_{pk} = \pm f_{s-p+1,k};, ~\forall k$. The choice of the sign depends on the index $k$. Consequently, we split the sums in (\ref{ap}) according to the parity of the eigenstates. To do so, we divide all the possible indexes $k,l$ in four sets $\{\Lambda_{++},\Lambda_{-+},\Lambda_{+-},\Lambda_{--}\}$ such that if $k,l \in \Lambda_{ab}$ then $f_{p,k} = a f_{s-p+1,k}$ and $f_{p,l} =b f_{s-p+1,l}, \ a,b = +,- $.

One can also show that the spin inversion symmetry induces the relation $\delta_{p(j),\uparrow}=\delta_{(s-p+1)(j),\downarrow} ~\forall p$. As a consequence, we can write Eq.(\ref{mj}) as in the following:
\begin{equation}
m_j(t) = 4 \sum_{p=1}^{s/2} \sum_{l,k \in \Lambda_{+-}} \cos(\omega_{k,l}t) f_{p,k}f_{p,l}\psi_{l0}\psi_{k0} (\delta_{p(j),\uparrow} - \delta_{p(j),\downarrow} ).
\label{this-eq-2}
\end{equation}
From Eq. (\ref{this-eq-2}) we see that the frequencies $\omega_{k,l} = (E_k - E_l)/\hbar$, that have a non-vanishing contribution in the magnetization dynamics, are the ones connecting eigenstates with different parity. As immediate remark, we see that the frequencies $\omega_{k,k}=0$ don't affect the time evolution consistently with the Fourier analysis presented in Fig.~3c. 

\begin{figure}
\includegraphics[width=1\textwidth]{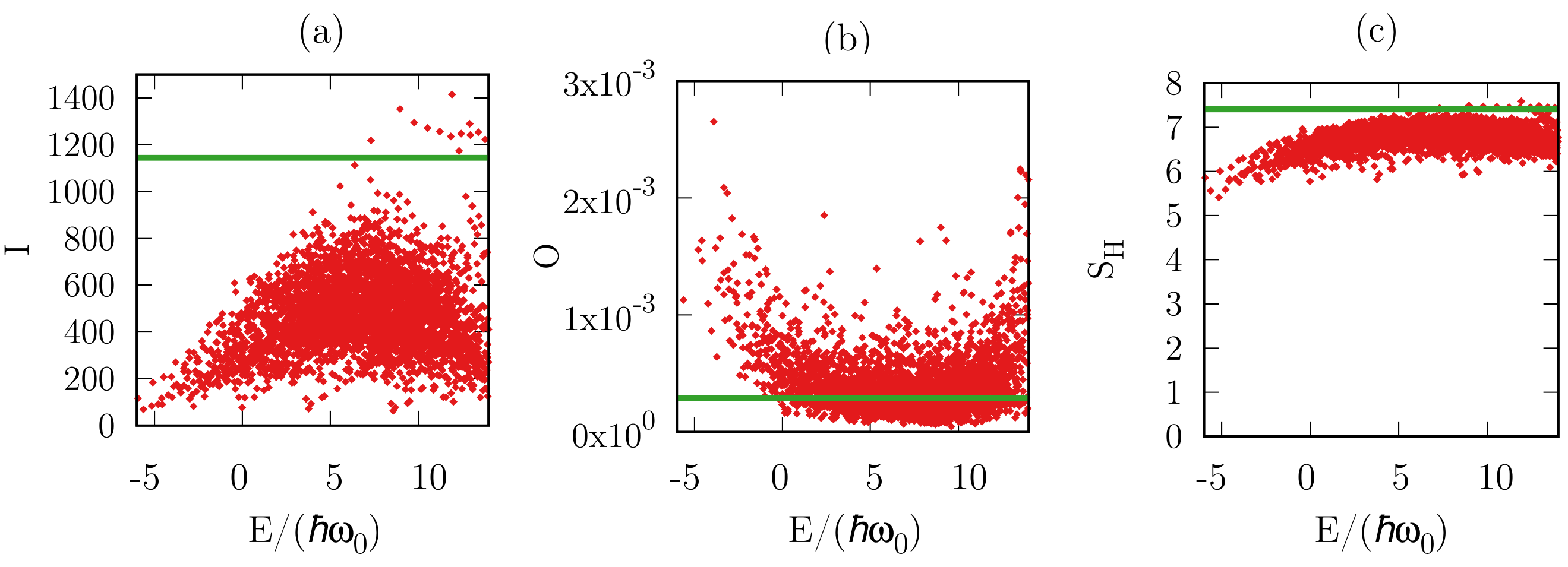}
\caption{From left to right respectively: (a) inverse partecipation ratio $I_r$ (b) overlap between the probability distributions of neighboring eigenstates $O$ and (c) Shannon entropy $S_H$ as functions of the energy, in units of $\hbar \omega_0$, for a system of $N=14$ particles. The green lines indicate the corresponding value predicted by Random Matrix Theory.}
\label{AppendixC}
\end{figure}

\subsection{Larger systems}
The results presented in the main text are obtained using exact techniques. Such approach yields a deep understanding of the physical phenomena involved, however, the exact solution gets more and more complex as the number of degrees of freedom increases. 

In order to overstep such limitation and to corroborate the results presented in the main text, we have performed numerical simulations  using the time-dependent Density-Matrix-Renormalization Group (DMRG) method \cite{PhysRevLett.93.076401,schollwock2006methods} . In particular, in order to describe the long-time dynamics, we implemented an approximation of the time evolution based on the Taylor expansion of the evolution operator \cite{al2011computing}.

For the short time dynamics, the DMRG results for particle numbers $N=20$ to $N=60$ are presented in Fig.$2$ of the main text. In Fig.~\ref{AppendixD} we show the results for intermediate times. We first of all 
compare the time evolution of the discrete center of mass of the magnetization $d_D(t) = \sum_j j \ m_j(t)$ with the corresponding quantity calculated exactly  in the continuum case $d(t)$ (panel a). We find that the two curves are proportional one to the other, \textit{i.e.} $d(t) = A(N)d_D(t)$. The constant $A(N)$ depends on the details of the harmonic trap and scales as $\propto N^{-1/2}$ (see inset in the panel a) of Fig.~\ref{AppendixD}). In panel (b) of the same figure we show  $d_D(t)$ for different number of particles, as a function of the rescaled time $\omega_0 t/N^{1/4}$.  Apart from a small variation of the details of the oscillations, due to the higher number of modes involved in the dynamics, we find that  curves collapse onto each other, thus confirming that  the universal scaling reported in the main text holds also for larger values of $N$.

\subsection{Further probes of chaoticity}
We provide here further arguments on the breaking of the integrability of the model induced by the strong coupling expansion to first order in $1/g$. In the spirit of Ref.~\cite{PhysRevE.85.036209}, we calculated other quantities to probe the presence of chaoticity in the system. In particular,  we focused on the eigenstates of the Hamiltonian and on their localization properties. In Fig.~\ref{AppendixC} we show from left to right respectively the inverse participation ratio $I = \sum_{p=1}^s \abs{f_{p,j}}^{-4}$, the overlap between consecutive probability distribution, $O = \sum_{p=1}^{s-1} \abs{f_{p,j}}^2 \abs{f_{p,j+1}}^2$ and the Shannon entropy $S_H = - \sum_{p=1}^s \abs{f_{p,j}}^2 \ln \abs{f_{p,j}}^2 $. We compare the three quantities with the corresponding value predicted by the Random Matrix Theory, indicated by the green lines. By comparing the results with the ones of Ref.~\cite{PhysRevE.85.036209}, we see that the eigenstates shows some similarities in terms of statistical properties, but are not completely sparse as they would be in the integrable system.

\end{document}